# Superconductivity at 3.85 K in $BaPd_2As_2$ with the $ThCr_2Si_2$-type structure

Qi Guo[1], Jia Yu[1], Bin-Bin Ruan[1], Dong-Yun Chen[1], Xiao-Chuan Wang[1], Qing-Ge Mu[1], Bo-Jin Pan[1], Gen-Fu Chen[1,2], Zhi-An Ren[1,2] *

[1] Institute of Physics and Beijing National Laboratory for Condensed Matter Physics, Chinese Academy of Sciences, Beijing 100190, China

[2] Collaborative Innovation Center of Quantum Matter, Beijing 100190, China

* Email: renzhian@iphy.ac.cn

**Abstract**

The single crystal of $ThCr_2Si_2$-type $BaPd_2As_2$ was successfully prepared by a self-flux method, and the crystal structure was characterized by X-ray diffraction method with the lattice parameters a = 4.489(2) Å and c = 10.322(3) Å. Bulk superconductivity below a $T_c$ (critical temperature) of 3.85 K was revealed in this compound by the measurements of resistivity, magnetic susceptibility and specific heat, and this $T_c$ is much higher than those of other isostructural Pd-based superconductors.

1. **Introduction**

Transition metal compounds have been widely studied due to their intriguing physical properties which usually originate from the strongly correlated d-electrons, especially for the search of high-$T_c$ ($T_c$ is critical temperature) superconductors [1-3]. Among the new high-$T_c$ family of iron-pnictide superconductors, the ternary parent compounds $AFe_2As_2$ (A = Ca, Sr, Ba, *etc.*) belong to a common $ThCr_2Si_2$-type crystal structure, and they are spin density wave (SDW) semimetals at low temperature. Superconductivity can be induced after suppressing the SDW order by chemical doping, external pressure or internal crystal defects, for instance the $Ba_{1-x}K_xFe_2As_2$ superconducts at 38 K [4-6]. Many recent works were focused on these 122-type ternary superconductors to solve the hidden pairing mechanism in these high-$T_c$ materials, mostly because that high quality single crystal samples can be easily synthesized for these inter-metallic compounds [7-10]. Moreover, most of the transition metals can form this simple body-centered tetragonal $ThCr_2Si_2$-type crystal structure, and to study similar materials within this structure may be an easy route for the search of new superconductors. Some of these iron-free compounds were already reported to be superconductors, such as $SrNi_2As_2$, $LaRu_2P_2$, $BaRh_2P_2$, $SrPd_2Ge_2$, $SrIr_2As_2$, *etc.*, while their superconducting $T_c$ is all very low within several degrees Kelvin [11-14].

Recently, Anand *et al.* prepared single crystals of $APd_2As_2$ (A = Ca, Sr, Ba) by PdAs flux [15], and reported superconductivity for $CaPd_2As_2$ and $SrPd_2As_2$ which both crystallizes in the collapsed $ThCr_2Si_2$-type structure; while $BaPd_2As_2$ crystallizes in the tetragonal $CeMg_2Si_2$-type structure without bulk superconductivity. The transition metal Pd-containing $ThCr_2Si_2$ type compounds have been studied by several groups, and some discovered superconductors are $LaPd_2Ge_2$ ($T_c$ = 1.12 K) [16], $CaPd_2Ge_2$ ($T_c$ = 1.98 K) [17], $SrPd_2Ge_2$ ($T_c$ = 3.04 K) [14], $CaPd_2As_2$ ($T_c$ = 1.27 K), $SrPd_2As_2$ ($T_c$ = 0.92 K) [15], and $LaPd_2As_2$ ($T_c$ = 1 K) *etc.* [18]. Another 122-type $LaPd_2Sb_2$ ($T_c$ = 1.4 K) superconductor forms in a $CaBe_2Ge_2$-type structure [19]. We have carefully studied the Ba-Pd-As ternary phases, and prepared three polymorphs of the $BaPd_2As_2$ compounds, namely the $CeMg_2Si_2$-type [20], the $BaPd_2As_2$-type [20], and the ThCr$_2$Si$_2$-type [21], and found superconductivity only exists in the last structure with measurements above 1.8 K. Here we report the preparation of the ThCr$_2$Si$_2$-type BaPd$_2$As$_2$ single crystal, and bulk superconductivity was revealed in it by resistivity, magnetization and specific heat measurements with a $T_c$ of 3.85 K, which is interestingly much higher than all other 122-type Pd-based superconductors.

**2. Experimental details**

The single crystal of BaPd$_2$As$_2$ samples were grown by a self-flux method. The starting materials of high-purity elemental Ba pieces, Pd and As powders were mixed together by a molar ratio of 1:2:2 (as stoichiometric BaPd$_2$As$_2$), then placed into alumina crucibles and sealed in evacuated quartz tubes. The sealed tubes were slowly heated up to 1100 °C at a rate of 50 °C/h and kept at this temperature for 20 hours, then slowly cooled down to 800 °C at a rate of 2 °C/h. The furnace was then cooled down to room temperature. Shiny millimeter-size plate-like crystals were easily separated. We note that the molar ratio of the starting elements is the most crucial factor for the successful growth of ThCr$_2$Si$_2$-type BaPd$_2$As$_2$ single crystals.

The crystal structure of all samples was characterized by powder X-ray diffraction (XRD) method on a PAN-analytical x-ray diffractometer with Cu-K$_\alpha$ radiation at room temperature. The resistivity and specific heat were measured by a Quantum Design physical property measurement system (PPMS), and the magnetic susceptibility was measured with both field cooling (FC) and zero field cooling (ZFC) methods under a magnetic field of 5 Oe by a Quantum Design magnetic property measurement system (MPMS).

**3. Results and discussion**

The XRD patterns for the powders of crushed single crystals and the surface of a piece of single crystal were collected and presented in Fig. 1. All the diffraction peaks of the powder XRD patterns in Fig. 1(a) can be well indexed with the *I4/mmm* (No. 139) space group of the ThCr$_2$Si$_2$-type crystal structure, as depicted in the inset of Fig. 1(a). The refined lattice parameters are a = 4.489(2) Å and c = 10.322(3) Å, similar to those previously reported [21]. No CeMg$_2$Si$_2$-type BaPd$_2$As$_2$ crystal was found in this batch of samples. The optical image of a piece of as-prepared single crystal is shown in Fig. 1(b) with a typical size of 3×2×0.2 mm$^3$, showing a layered crystal structure. The XRD patterns for the surface of the single crystal in Fig. 1(b) only show the (00$l$) ($l = 2n$) diffraction peaks, suggesting the good uniformity of the crystal with the surface perpendicular to the $c$ axis. The chemical composition analysis was performed by the energy-dispersive spectroscopy (EDS) method for several pieces of crystals, and the results indicate that the average atomic ratio is close to Ba:Pd:As = 1:2:2, consistent with the 122-type BaPd$_2$As$_2$ constituent. During our crystal growth, if more arsenic was added in the flux, only CeMg$_2$Si$_2$-type BaPd$_2$As$_2$ crystals were obtained.

The temperature dependence of in-plane electrical resistivity for the BaPd$_2$As$_2$ crystal was measured from 1.8 K to 300 K, which is shown in Fig. 2(a). A very sharp superconducting transition is observed when the temperature down to a $T_c$(onset) of 3.85 K, and the $T_c$(zero) is ~ 3.80 K where resistivity reaches 0. This $T_c$ is much higher than those of its superconducting analogs of CaPd$_2$As$_2$ and SrPd$_2$As$_2$ *etc.*. The reason that leads to a higher $T_c$ in BaPd$_2$As$_2$ deserves further studies. Above $T_c$, the resistivity data display metallic behavior similar to its analogs, but with a relatively higher residual resistivity ratio $RRR = \rho(300K) / \rho(4K) = 12.5$, which shows the good crystal quality of our samples [15]. The superconducting transition was also characterized under variable magnetic fields as shown in Fig. 2(b). The $T_c$ decreases almost linearly with the applied fields, and superconductivity disappears at 1.8 K under a field of 0.14 T. The upper critical field $\mu_0H_{c2}$(T) was estimated from the mid-point of the resistivity transition with a linear extrapolation, which gives a small value of $\mu_0H_{c2}$(0) about 0.21 T, that is close to the CaPd$_2$As$_2$ and SrPd$_2$As$_2$ superconductors.

In Fig. 3, the temperature dependence of DC magnetic susceptibility was measured between 1.8 K and 300 K with both zero-field-cooling (ZFC) and field-cooling (FC) methods under a field of 5 Oe. The onset transition of the Meissner effect also appears at 3.85 K, with a large superconducting shielding volume fraction about 87% at 1.8 K. The susceptibility data reveal the typical type-II bulk superconductivity feature in the ThCr$_2$Si$_2$-type BaPd$_2$As$_2$ compound. No superconducting trace was found in CeMg$_2$Si$_2$-type BaPd$_2$As$_2$ crystals in our experiments as mentioned in Ref. [15]. Above the superconducting transition, the susceptibility shows weak diamagnetic behavior, and no other magnetic order is observed.

The low temperature specific heat $C_p$ was also measured to further confirm the bulk superconductivity in BaPd$_2$As$_2$ crystal. The data for $C_p/T$ versus $T^2$ are plotted in Fig. 4(a) from 2 K to 5 K under the field of 0 T and 1 T respectively. A clear specific heat jump happening at 3.8 K supports the bulk nature of superconductivity, and this jump vanishes under the field of 1 T which exceeds the superconducting upper critical field. At normal state, the data of $C_p$ (with and without field) coincide with each other very well, and these data were fitted by the formula $C_p/T = \gamma + \beta T^2$ with 15 K$^2$ < $T^2$ < 25 K$^2$. The data at lower temperature was not fitted due to the unusual decrease from the linearity, which need further study with ultra-low temperature measurements. The fitted values of the Sommerfeld coefficient $\gamma$ and $\beta$ are 12.1(1) mJ/mol K$^2$ and 3.77(8) mJ/mol K$^4$ respectively, and the $\beta$ deduces a Debye temperature $\Theta_D$ of 137 K. The electronic contribution of specific heat $C_e$ without field was obtained by subtracting the phonon contribution from $C_p$, and the curve $C_e/T$ versus $T$ is plotted in Fig. 4(b). The specific heat jump determined from the data at $T_c$ yields the $\Delta C_e/\gamma T_c$ = 1.41, which is close to the BCS value of 1.43 in the weak coupling limit.

In conclusion, we have successfully grown the high quality BaPd$_2$As$_2$ single crystal with ThCr$_2$Si$_2$-type structure, which is a type-II bulk superconductor with a $T_c$ of 3.85 K.

**Acknowledgments**

The authors are grateful for the financial supports from the National Natural Science Foundation of China (No. 11474339), the National Basic Research Program of China (973 Program, No. 2010CB923000 and 2011CBA00100) and the Strategic Priority Research Program of the Chinese Academy of Sciences (No. XDB07020100).

**Figure Captions:**

Figure 1: XRD patterns with indexed diffraction peaks for BaPd$_2$As$_2$ samples of (a) the powders of crushed single crystals and (b) the surface of a piece of single crystal, the insets show the crystal structure and a piece of crystal sample.

Figure 2: (a) The temperature dependence of in-plane electrical resistivity for the BaPd$_2$As$_2$ crystal. (b) Expanded plots of the resistivity transitions under variable magnetic fields. The inset in (a) shows the linear extrapolation of the $\mu_0H_{c2}$(T).

Figure 3: The temperature dependence of FC and ZFC magnetic susceptibility for the BaPd$_2$As$_2$ crystal under a field of 5 Oe. The inset shows the expanded curve at the transition.

Figure 4: (a) The specific heat $C_p/T$ versus $T^2$ for 2 K < $T$ < 5 K under the fields of 0 T and 1 T for the BaPd$_2$As$_2$ crystal, with a linear fit (the dotted line) for the normal state. (b) The electronic contribution of specific heat $C_e/T$ versus $T$, which indicates the electronic specific jump at $T_c$ and the Sommerfeld coefficient $\gamma$ (the intercept of the horizontal dotted line).

**Fig. 1.**

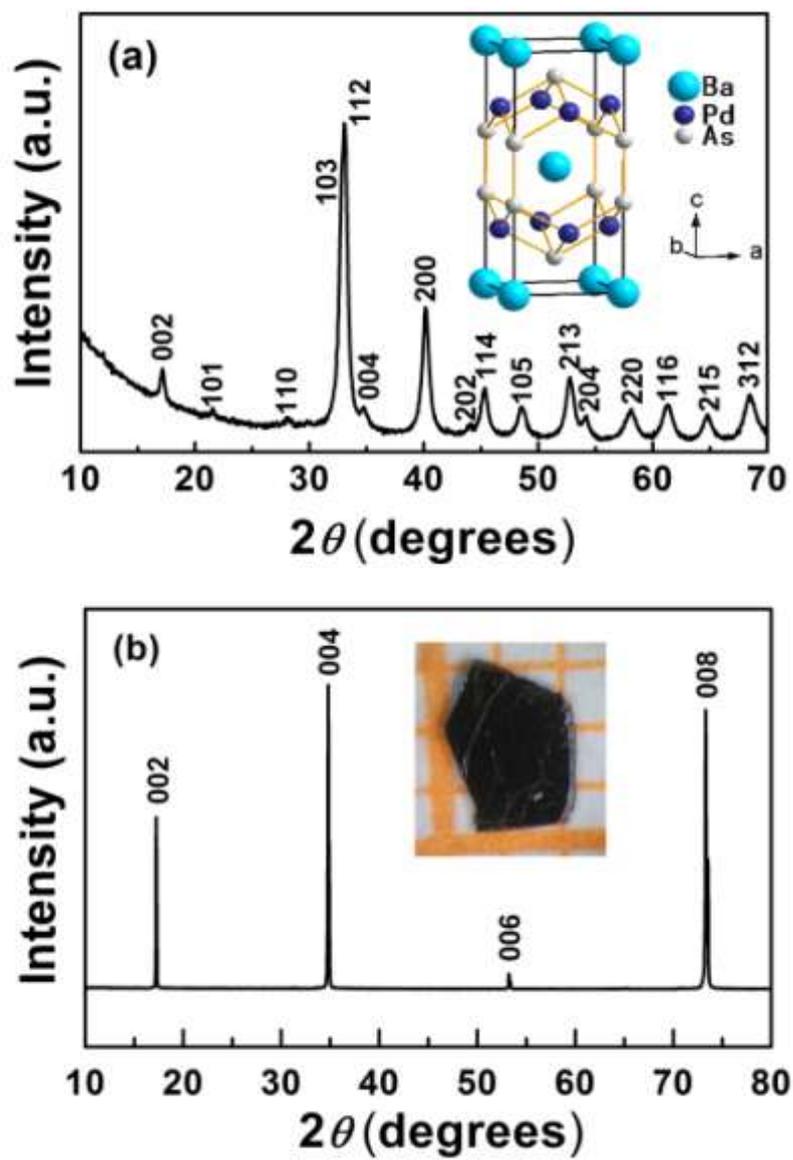

**Fig. 2.**

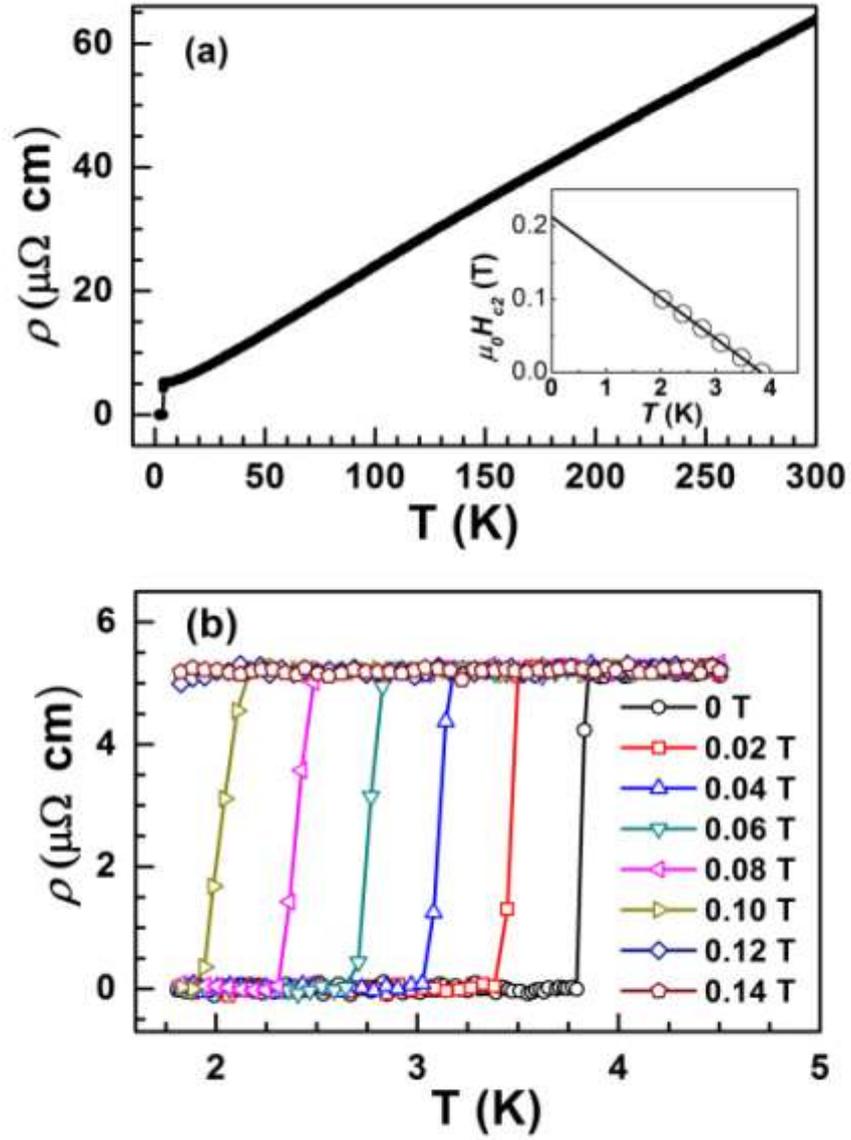

**Fig. 3.**

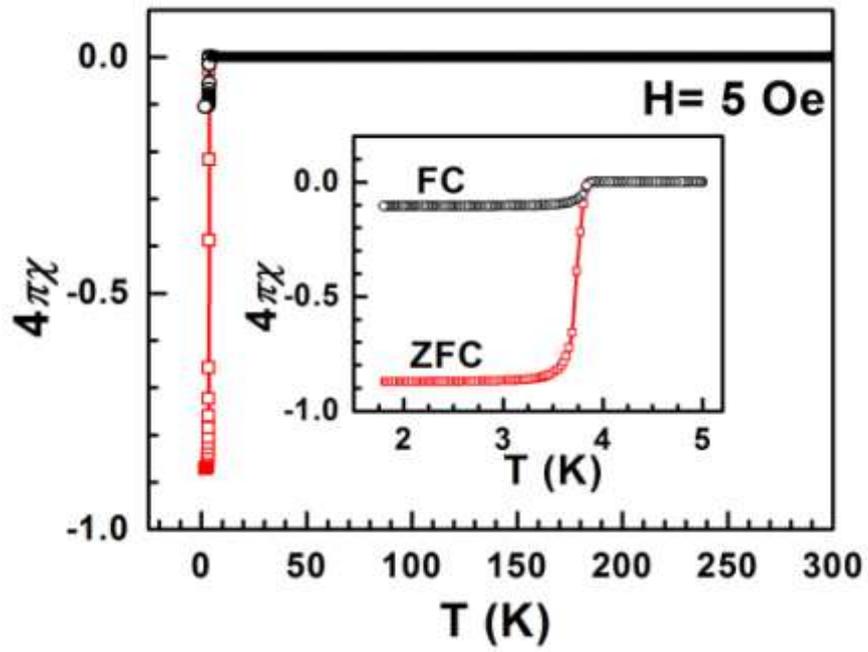

**Fig. 4.**

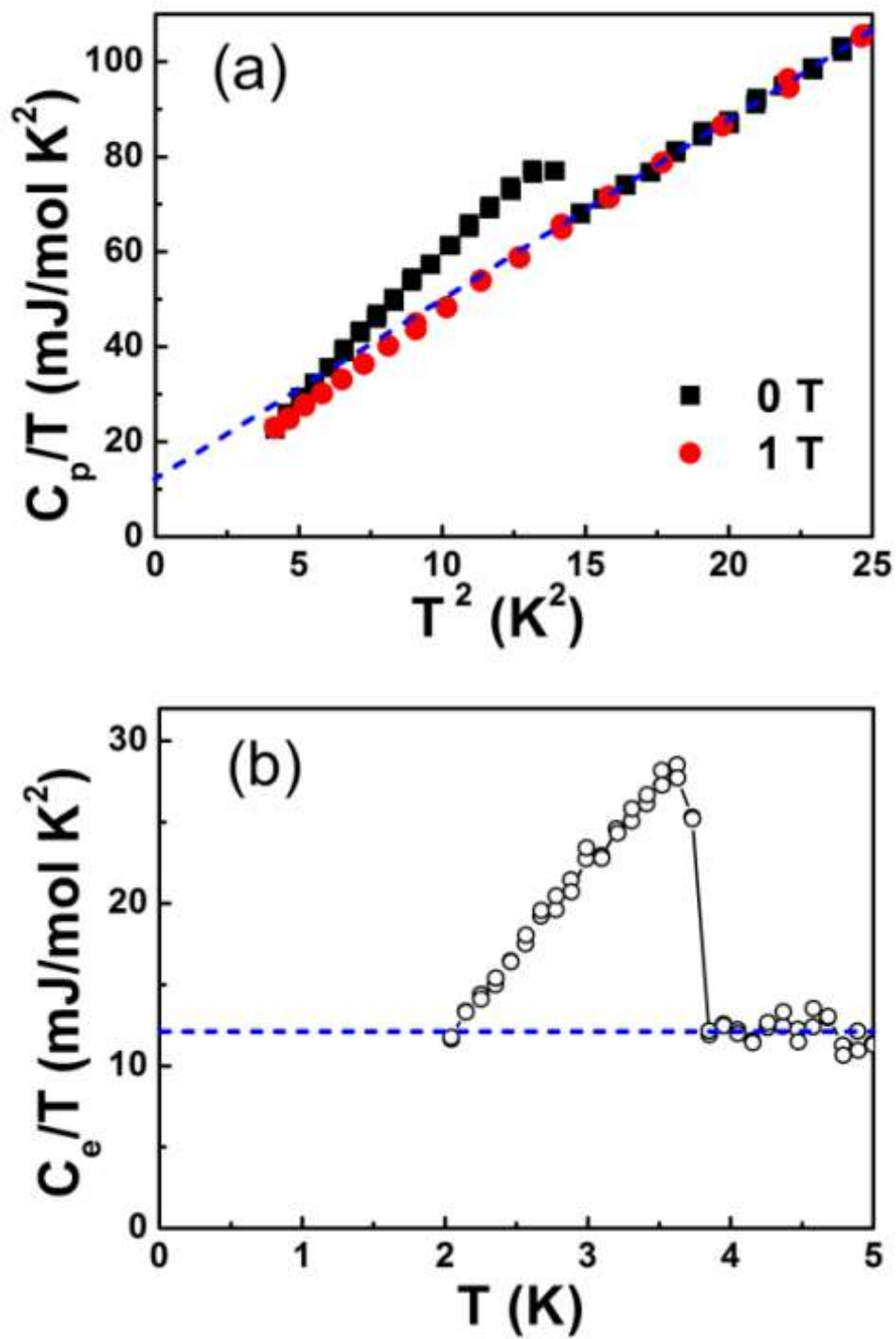